\documentstyle[psfig]{mn}
\newcommand{\reference}{\bibitem}
\def\beq{\begin{equation}}
\def\eeq{\end{equation}}
\def\cm{\,{\rm {cm}}}

\def\kpc{\,{\rm {kpc}}}

\def\kms{\,{\rm {km\, s^{-1}}}}
\def\erg{\,{\rm erg}}
\def\msun{{\rm M}_\odot}
\def\vcir{V_{\rm c}}

\def\Obaryon{{\Omega_{\rm B,0}}}
\def\Kdegree{\,{\rm K}}
\def\kkev{\,{\rm keV}}
\def\kev{\,{\rm keV}}
\def\keV{\,{\rm keV}}
\def\yr{\,{\rm yr}}
\def\sec{\,{\rm s}}

\def\v200{{V_{200}}}
\def\r200{r_{\rm vir}}
\def\M200{M_{\rm vir}}
\def\s100{{\cal S}_{100}}

\def\Tvir{T_{\rm vir}}
\def\jvir{j_{\rm vir}}

\def\e0{{\epsilon_0}}

\def\rmd{{\rm d}} 

\def\Lya{{\rm Ly}\alpha}
\def\mui{{\mu_{\rm i}}}

\input{psfig.sty} 
\begin{document}
\title[Galaxy Formation in Preheated Intergalactic Media]	
{Galaxy Formation in Preheated Intergalactic Media}
\author[]	
{H.J. Mo$^1$, Shude Mao$^{2}$
\thanks {E-mail: hom@mpa-garching.mpg.de, smao@jb.man.ac.uk} 
\\
\smallskip	
      $^1$Max-Planck-Institut f\"ur Astrophysik
      Karl-Schwarzschild-Strasse 1, 85748 Garching, Germany \\
      $^2$Jodrell Bank Observatory, Macclesfield, Cheshire SK11 9DL, UK
}
\date{Accepted ........
      Received .......;
      in original form .......}
\maketitle

\begin{abstract}
 We outline a scenario of galaxy formation in which   
the gas in galaxy-forming regions was preheated 
to high entropy by vigorous energy feedback 
associated with the formation of stars
in old ellipticals and bulges and with AGN activity. 
Such preheating likely occurred at redshifts $z\sim 2$ -- 3, 
and can produce the entropy excess 
observed today in low-mass clusters of galaxies 
without destroying the bulk of the ${\rm Ly}\alpha$ forest.
Subsequent galaxy formation is affected by the preheating,
because the gas no longer follows the dark matter on 
galaxy scales. The hot gas around galaxy haloes has 
very shallow profiles and emits only weakly in the X-ray.  
Cooling in a preheated halo is not inside-out,
because the cooling efficiency does not change 
significantly with radius. Only part of the gas in a
protogalaxy region can cool and be accreted 
into the final galaxy halo by the present time. The accreted gas 
is likely in diffuse clouds and so does not lose 
angular momentum to the dark matter. Cluster ellipticals 
are produced by mergers of stellar 
systems formed prior to the preheating, while
large galaxy disks form in low-density environments 
where gas accretion can continue to the present time. 
\end{abstract}

\section{INTRODUCTION}
\label{sec1}

 In the past 20 years, the Cold Dark Matter (CDM) cosmogony 
(Peebles 1982; Blumenthal et al. 1984) has become
the most appealing scenario for the formation of structure in the universe. 
In the CDM cosmogony, the universe is assumed to be dominated by 
CDM, and the structure in the universe forms in a hierarchical 
fashion, with dark matter particles aggregating into larger 
and larger clumps (dark haloes) in the passage of time. 
In this scenario, galaxies are assumed to form as gas cools 
and condenses in dark haloes (White \& Rees 1978),
and the properties of galaxies are determined by the properties
of their host haloes, such as mass, density profile, 
angular momentum, and merging history. This standard scenario
of galaxy formation has been quite successful 
in predicting many of the observed properties of the 
galaxy population, such as disk sizes and kinematics
(e.g. Fall \& Efstathiou 1980; Dalcanton, Spergel \& Summers
1997; Mo, Mao \& White 1998, hereafter MMW; 
Avila-Reese, Firmani \& Hernandez 1998; 
Heavens \& Jimenez 1999; Mo \& Mao 2000; 
Navarro \& Steinmetz 2000; van den Bosch 2000), 
and the distributions of luminosity, colour and morphology 
(e.g. White \& Frenk 1991; Kauffmann, White \& Guiderdoni 1993; 
Cole et al. 1994; Somerville \& Primack 1999).

One key ingredient in the standard scenario is 
a mechanism which can prevent gas from cooling 
too fast in dark matter haloes, because without such 
a mechanism, we may run into the following set of problems.
\begin{itemize}
\item 
The overcooling problem: In a hierarchical model, all dark
matter was in small haloes at high redshifts. Since radiative
cooling of gas in these haloes is effective, all the gas 
would have cooled and formed stars, leaving no gas for 
producing quasar absorption line systems and for the formation 
of galaxy disks at low redshifts. 
This is the `overcooling problem' first pointed out by 
White \& Rees (1978).  
\item
The baryon-fraction problem: Based on cosmic nucleosynthesis 
(e.g. O'Meara et al. 2001) and Cosmic Microwave Background 
Radiation (e.g. de Bernardis et al. 2002),
the baryon density in the universe is $\Obaryon\approx 0.02 h^{-2}$. 
For a universe with $\Omega_0\approx 0.3$ and $h\approx 0.7$, 
the implied overall baryon fraction is about 
$f_{\rm B}\equiv\Obaryon/\Omega_0\approx 0.13$, 
consistent with the baryon fraction in rich clusters
(e.g. White et al. 1993; Ettori \& Fabian 1999). 
On the other hand, modelling of galaxy disk formation 
in CDM haloes showed that the gas mass 
which settles into the disk is only $\sim 5\%$ of the 
total halo mass and a much higher fraction would 
produce disks which might be unstable and have too peaked 
rotation curves (e.g. MMW). 
Since the total amount of hot gas in a galaxy halo is small,
a consistent model of disk formation requires that
only part of the gas in a protogalaxy region ends up in the
final halo. The question is then what regulates the 
baryon fraction in galaxy haloes.
\item
Angular-momentum problem: Since gas can cool rapidly in small
progenitor haloes to form condensed central objects, it can 
lose most of its angular momentum to the dark matter during 
galaxy assembly (e.g. Navarro \& White 1994).
This leads to serious problems when comparing with real 
galaxies: the disks formed in this way are too small.
\item
 X-ray halo problem: In the standard scenario, gas in a galactic 
halo is assumed to be shock-heated to the virial temperature
(which is about $10^6\Kdegree$) by gravitational collapse, and
(massive) galactic haloes are predicted to emit in X-ray, 
with a bolometric luminosity of the order $10^{42}\erg\sec^{-1}$. 
The predicted luminosity is about one order of magnitude higher 
than what is observed in the haloes of spiral galaxies 
(e.g. Benson et al. 2000). This is a well-known problem 
(e.g. White \& Frenk 1991) but has been largely ignored in 
theoretical studies of galaxy formation so far.   
\end{itemize}

Two mechanisms have been proposed to solve 
these problems. The first is photoionization heating 
of the interglactic medium (IGM) by the UV background. 
Detailed analyses showed that 
photoionization heating is effective in prohibiting 
gas cooling only in small haloes, with circular 
velocity $\vcir\la 50\kms$ (e.g. Efstathiou 1992; 
Gnedin 2000). Thus, while photoionization heating may be 
able to keep some of the IGM in a diffuse form
(and so alleviating the overcooling problem), its effect
on gas assembling and cooling in galactic haloes may
not be sufficient to solve the other problems mentioned.
The second possibility is heating by the energy feedback 
from star formation (Larson 1974; White \& Rees 1978; 
Dekel \& Silk 1986). In many semi-analytic models of 
galaxy formation (e.g. White \& Frenk 1991; 
Kauffmann et al. 1993; Cole et al. 1994; 
Somerville \& Primack 1999; Kauffmann et al. 1999), 
this feedback is implemented in such a way that the
gas in galactic haloes is assumed to be effectively 
heated by supernova explosions associated with the 
star formation in the galaxy, and the heated gas is 
assumed either to be retained within the halo or 
ejected from the halo to some larger scale. 
Detailed modelling showed that the total energy feedback 
from supernova explosions in a galaxy is sufficient to 
keep a large amount of gas at a high temperature, and 
the feedback efficiency can be tuned to reproduce
many of the observed properties of the galaxy population.
Unfortunately, our understanding of such feedback is still 
very limited, and it is unclear how it operates in detail.

 In this paper, we consider another possible 
implementation of the feedback process. 
We assume that the IGM in galaxy-forming regions 
was heated to some high entropy at some high redshift 
during an early period of vigorous star formation and 
associated AGN activity. We examine the effect of such 
preheating on subsequent galaxy formation. 
This implementation is motivated by three lines of 
observational evidence.
(i) Observations show that many stars in the universe 
may have formed at $z\ga 2$ in systems reminiscent of 
local starburst galaxies (see Heckman 2001 and references therein). 
Such galaxies can produce supernova-driven outflows which may be able to 
heat the IGM in the surrounding regions.
(ii) The comoving number density of AGNs is highest 
around redshifts $z=2$ - 3 (Shaver et al. 1996). Since a lot of 
kinetic energy can be released from AGNs, the surrounding 
IGM can be heated to a high temperature, if some of this kinetic
energy is thermalized (e.g. Inoue \& Sasaki 2001).
(iii) Observations of X-ray clusters and groups suggest 
that there is an entropy excess in these systems relative
to that expected from gravitational collapse 
alone (e.g. Ponman, Cannon \& Navarro 1999; 
Lloyd-Davies et al. 2000). 
This can be explained if the IGM in protocluster
(protogroup) regions was preheated to a high temperature 
(e.g. Ponman et al. 1999; Tozzi \& Norman 2001). 
The implementation considered here is related to some of 
the early implementations (e.g. Nulsen \& Fabian 1995, 1997;
Sommer-Larsen, Gelato \& Vedel 1999), 
but has some distinct properties. 
We emphasize the importance of early starbursts 
and associated AGN activity, and we consider heating of the gas 
while it was still in diffuse form outside galactic haloes.     
Furthermore, our model presents more detailed discussions 
about the properties of the preheated gas
in connection to the puzzling features of observed 
galaxies mentioned above.  
 
 Preheating affects subsequent galaxy formation because 
the heated gas cannot collapse into small dark 
haloes and so no longer follows the bottom-up clustering 
hierarchy of the dark-matter component. 
We show that the introduction of preheating 
can help to solve some of the problems mentioned 
above. The paper is structured as follows.
In Section \ref{sec2}, we describe our assumptions on 
the properties of the preheated IGM and analyze
gas cooling and accretion onto dark haloes in such a 
medium. The formation of galaxies in the present scenario is 
described in Section \ref{sec3}. In Section \ref{sec4},
we discuss briefly how the IGM may have been preheated. 
Further discussion of the 
present model and a summary of our main conclusions
are given in Section \ref{sec5}. Throughout this paper, 
when necessary, we 
adopt the standard $\Lambda$CDM cosmogony,
with matter density $\Omega_0=0.3$, a cosmological
constant corresponding to $\Omega_\Lambda=0.7$, and
a normalization $\sigma_8=0.9$ for the power spectrum.
We write the Hubble constant as 
$H_0=100h\,{\rm km\,s^{-1}Mpc^{-1}}$ and
take $h=0.7$.

\section {Gas accretion by dark haloes in a 
preheated intergalactic medium}
\label{sec2}

\subsection {The preheated IGM}

For a (completely ionized) cosmic gas, the mean number 
density of electrons at redshift $z$ is
\begin{equation}
{\overline n}_e\approx 2.0\times 10^{-7} 
\left({\Obaryon h^2\over 0.02}\right) (1+z)^3 
\cm^{-3}\,,
\end{equation}
where $\Obaryon$ is the cosmic density parameter of 
baryons, and $\Obaryon h^2\sim 0.02$ as given by 
cosmic nucleosynthesis (e.g. O'Meara et al. 2001).
Now consider a volume in which the
electron density is $n_e={\overline n}_e (1+\delta)$
and the temperature is $T$. We define an entropy parameter as
\begin{equation}
{\cal S}\equiv {T\over n_e^{2/3}}\,, 
\end{equation}
where it is conventional to express the temperature in units of keV 
($1\kkev$ corresponds to $T \approx 1.2\times 10^7\Kdegree$).
If the gas is adiabatic, then ${\cal S}$ is conserved and 
the temperature of the gas can be written in terms 
of ${\cal S}$ as 
\begin{equation}\label{temp_s100}
T=4.4\times 10^4 \s100 (1+\delta)^{2/3}
(1+z)^2\Kdegree\,,
\end{equation}
where 
\begin{equation}
\s100\equiv 
\left({ {\cal S} \over 100 h^{-1/3}\kkev\cm^2}\right)
\left({\Obaryon h^2\over 0.02}\right)^{2/3}
h_{0.7}^{-1/3},
\end{equation}
$h_{0.7}\equiv h/0.7$. 
Note that the $h$ dependence at the end of the equation 
appears because the observed value of ${\cal S}$ 
has such a dependence. 
For a given ${\cal S}$ the required 
temperature is lower if the gas density is 
lower. For example, if $\s100=1$ and if preheating 
was at $z=3$ in the mean medium ($\delta=0$), 
the required temperature would be 
$T\sim 7.5\times 10^5\Kdegree$. If, on the other hand, 
heating is in the central region of a present-day cluster,
where $\delta\sim 10^4$, the required temperature 
would be $T\sim 2\times 10^7\Kdegree$.   
 
The time-scale of radiative cooling, defined as 
$t_{\rm cool}\equiv T/{\dot T}$, can be written as
\begin{eqnarray}
t_{\rm cool}
&\approx& {3 k T \mui \over 2 \mu n_e \Lambda (T)}\nonumber\\
&\sim& 
1.0\times 10^{11}
\Lambda_{-23}^{-1} T_6 \left({\Obaryon h^2\over 0.02}\right)^{-1}
\nonumber\\
&&\times
(1+\delta)^{-1}\left({1+z\over 4}\right)^{-3} \yr\,,
\end{eqnarray}
where $T_6\equiv T/10^6\Kdegree$, $\mu\approx 0.6$ and $\mui \approx 1.2$
are the mean molecular weights per
particle and per ion, respectively, and
$\Lambda_{-23}$ is the cooling function $\Lambda (T)$
in units of $10^{-23}\cm^{3} \erg\sec^{-1}$ as defined in
Sutherland \& Dopita (1993). Notice that $\mui$ enters because 
Sutherland \& Dopita (1993) defines the cooling rate per unit volume
as $n_e n_i \Lambda(T)$, where $n_i$ is the total ion density.
If we approximate the cooling 
function as a power law of gas temperature, 
$\Lambda (T)\propto T^{-\alpha}$, the cooling time  
in an approximately isentropic gas scales as
\beq
t_{\rm cool}\propto {T\over n_e \Lambda (T)}
\propto T^{\alpha-1/2}\propto n_e^{2(\alpha-1/2)/3}\,.
\eeq
Thus, depending on whether $\alpha>1/2$ or $\alpha<1/2$,   
the cooling time is longer or shorter for a high-density 
(high-temperature) gas. This is different from an 
isothermal gas, where $t_{\rm cool}$ is always shorter
for higher $n_e$. 

The cooling time given above should be compared to 
the Hubble time 
\begin{equation}
t_{\rm H}=3.2\times 10^9 h_{0.7}^{-1} 
\left({\Omega_0\over 0.3}\right)^{-1/2}
\left({1+z\over 4}\right)^{-3/2} {\cal H} (z)\,{\rm yr}\,,
\end{equation}
where 
${\cal H}(z) \equiv \Omega_{0}^{1/2} (1+z)^{3/2} H_0/H(z)$.
Note that ${\cal H}(z)\sim 1$ for $z\ga 1$ in a flat 
universe with $\Omega_0\sim 0.3$. 
As one can see, cooling is ineffective at
low overdensities and at low preheating redshift ($z\la 3$). 
In this case, the specific entropy
${\cal S}$ is approximately conserved (except in shocks),
and the evolution of the 
temperature is given by equation (\ref{temp_s100}). 
The properties of the preheated gas are then specified by
its density [which is proportional to $(1+\delta) (1+z)^3$] 
and the value of ${\cal S}$. 

One way to specify the initial entropy is to use the observed 
entropy excess in clusters and groups 
(Ponman, Cannon \& Navarro 1999; Lloyd-Davies et al. 2000), 
which gives $ {\cal S} \sim 100 \kev\cm^2$. We use this as the 
fiducial value for ${\cal S}$. It must be emphasized, however, 
that in reality the value of ${\cal S}$ may change from 
place to place, but we ignore this complication here. 

 In Section \ref{sec4}, we will discuss in some detail 
how the IGM may have been preheated. Based on the discussion 
there, we advocate a preheating process which can be 
summarized as follows:
\begin{itemize}
\item The preheating was due to the starbursts
(and associated AGN activity) 
responsible for the formation of old stars in 
ellipticals and bulges.  
\item  
The preheating was likely to be confined to protocluster
and protogroup regions where starbursts were common.
\item  
Although preheating must be a continuous process and
different regions may have been preheated at different 
times, the bulk of preheating was likely to be around 
a redshift $z\sim 2$ -- 3, where the star-formation 
rate and AGN number peak.  
\end{itemize}

The details of the preheating process are still poorly
understood, but much of the following discussion is 
independent of these details.

\subsection{Accretion of the hot IGM}

\begin{figure}
\centering  
\vskip-0.5cm
\psfig{figure=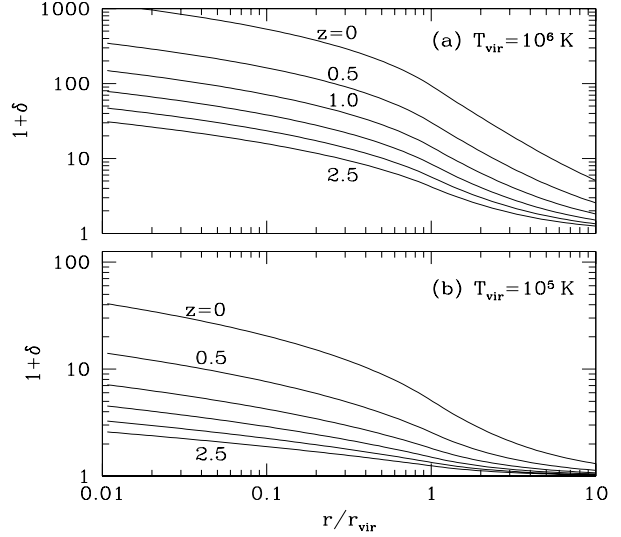,width=8.5cm,height=8.5cm,angle=0}
\vskip-0.5cm
\caption{Density profiles of preheated gas (assumed to be 
adiabatic, with $\s100=1$) as a function of radius
around dark haloes. Six curves are shown for
redshift from 0 to 2.5 with a step size of 0.5.
Results are shown 
for (a) a normal galaxy halo with $\Tvir=10^6\Kdegree$
and (b) a dwarf galaxy halo with $\Tvir=10^5\Kdegree$.}
\label{fig_density}
\end{figure}
\begin{figure}
\centering  
\vskip-0.5cm
\psfig{figure=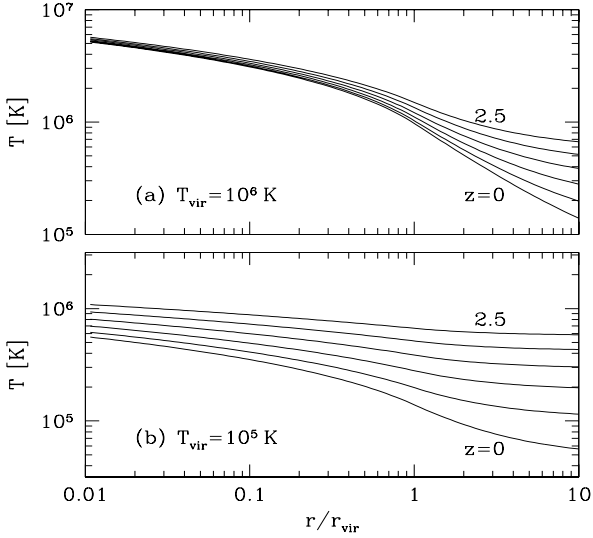,width=8.5cm,height=8.5cm,angle=0}
\vskip-0.5cm
\caption{Temperature profiles of preheated gas (assumed to be 
adiabatic, with $\s100=1$) as a function of radius
around dark haloes.  Results are shown 
for (a) a normal galaxy halo with $\Tvir=10^6\Kdegree$
and (b) a dwarf galaxy halo with $\Tvir=10^5\Kdegree$.}
\label{fig_temperature}
\end{figure}

Consider a dark halo with mass $\M200$, 
circular velocity $\vcir$, virial temperature $\Tvir$, and 
virial radius $\r200$. According to the spherical collapse model, 
these quantities are related as follows:
\beq \label{eq:virial}
\M200={\vcir^3\over 10G H(z)}\,,~~~
\r200={\vcir\over 10 H(z)}\,,~~~
\Tvir={\mu m_{\rm p} \vcir^2\over 2k}\,.
\eeq
The virial temperature is the same
as the isothermal temperature for a singular isothermal
sphere, but its meaning is less clear for more complex
halo profiles. It should only be regarded as an approximate
measure of the gas temperature in the halo.
We model the dark matter halo as 
a singular isothermal sphere with density profile
\beq
\rho(r)={\vcir^2\over 4\pi G r^2}\,.
\eeq
In a preheated medium where gas cooling is unimportant, 
the halo can accrete the preheated gas though adiabatic 
compression. If the gas accretion is subsonic
(Tozzi, Scharf \& Norman 2000), then we
may solve for the distribution of gas within the halo as follows.
Assuming spherical symmetry and hydrostatic equilibrium
(the latter assumption is valid in a region where the 
sound-crossing time is much shorter than the age of the universe), 
we have
\beq\label{hydrostatic}
{1\over\rho}{{\rm d} P\over {\rm d} r}=-{GM\over r^2}\,,
\eeq
where $\rho$ and $P$ are the density and pressure 
of the gas at radius $r$, $M$ is the mass within radius 
$r$. For the gas outside the halo, we may take $M$ to 
be the total mass of the halo. If the gas is not 
significantly shocked when establishing the hydrostatic
equilibrium, as is the case if the temperature of the 
preheated gas is not much lower than the virial 
temperature of the halo, the specific entropy $T/n_e^{2/3}$ will remain 
at the preheated value ${\cal S}$. In this case, 
the hydrostatic equilibrium 
equation can be solved for the gas density: 
\beq \label{deltaR0}
n^{2/3}={2\over 5}{G\M200 \mu m_p\over k {\cal S}}{1\over r}+{\rm constant}\,.
\eeq
Assuming that $n$ is equal to the mean density at $r\to \infty$, 
and using the relation between halo mass and halo virial 
temperature, we have 
\beq\label{deltaR1}
n^{2/3}(r)={\overline n}^{2/3}
+{4\over 5}{\Tvir\over {\cal S}}{\r200\over r},
~~~~ (r\ge \r200)\,.
\eeq
We caution that equation (\ref{deltaR1}) is only
approximate at large radii as the gas is likely not in 
hydrostatic equilibrium.
Within the virial radius of an isothermal sphere,
the hydrostatic equilibrium of adiabatic gas leads
to the following profile:
\beq\label{deltaR2}
n^{2/3}(r)={\overline n}^{2/3}
+{4\over 5}{\Tvir\over {\cal S}} \left(
1-\ln{r\over\r200} \right),
~~~~ (r<\r200)\,.
\eeq
The corresponding temperature profile is given by
\beq
T(r) = {\cal S} n^{2/3} (r)\,.
\eeq 
Some examples of the density and temperature
profiles are shown in Figures \ref{fig_density}
and \ref{fig_temperature}. 
As one can see, the density profile is very 
shallow. In fact this is the shallowest equilibrium 
profile one can get in a singular isothermal 
sphere, because any shallower profile is
convectively unstable. 

\subsection {X-ray emission from galactic haloes}

\begin{figure}
\centering  
\vskip-0.5cm
\psfig{figure=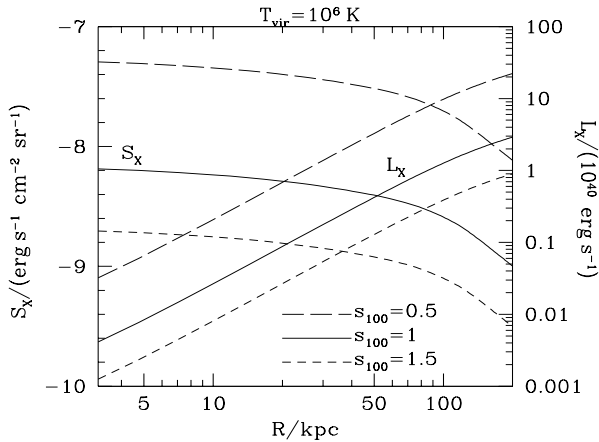,width=8.5cm,height=8.5cm,angle=0}
\vskip-1.5cm
\caption{The X-ray surface brightness $S_{\rm X}(R)$
and luminosity $L_{\rm X}(<R)$ (within radius $R$) 
of the hot gas around a present-day galactic halo with 
virial temperature $\Tvir=10^6\Kdegree$. Results 
are shown for three different values of $\s100$.}
\label{fig_Xray}
\end{figure}

 Given the density and temperature profiles, we
can estimate the radiation luminosity of the hot 
gas in a dark halo. Since the temperature of the 
gas is about $0.1\kev$ for a galaxy halo, 
the radiation is in the soft X-ray band, and we can hope to 
detect such emission through X-ray observations. 
The bolometric X-ray surface brightness is given by 
\beq
S_{\rm X}(R)={1\over 4\pi}\int\Lambda [T(r)] n_e n_i\,\rmd y,
\eeq
where $r=(R^2+y^2)^{1/2}$, 
and $R$ is the radius in projection.
The luminosity within a radius, $L_{\rm X}(<R)$,  
can then be obtained by integrating $S_{\rm X}$ 
over area and solid angle. 
Figure \ref{fig_Xray} shows 
$S_{\rm X}(R)$ and $L_{\rm X}(<R)$ for a present-day
galactic halo with virial temperature $\Tvir=10^6\Kdegree$,
corresponding to $\vcir\approx 170\kms$.
In the calculation, we have taken a metallicity $Z=0.1Z_\odot$ and 
neglected the contribution of the gas outside the virial
radius of the halo as the gas temperature becomes too low to be 
in the X-ray band (see Fig. 2).
For simplicity, we have also assumed that all the gas is 
in the hot phase. As one can see, the predicted 
X-ray emission depends significantly on the 
assumed specific entropy: a higher value of $\s100$
leads to a less concentrated gas distribution and 
so to lower X-ray surface brightness and luminosity.
For $\s100=1$, the surface brightness at 
$R=100$ and $20\kpc$ are $S_{\rm X}\sim 2.5\times 10^{-9}$ and
$5\times 10^{-9}\erg\sec^{-1}\cm^{-2}{\rm sr}^{-1}$,
respectively. The total luminosity within the virial
radius is $L_{\rm X}\sim 3\times 10^{40}\erg\sec^{-1}$.   
These predicted surface brightness and luminosity are about
a factor of 10 smaller than what one obtains based on
the assumption that gas in galactic haloes is
shock-heated to the virial temperature and in hydrostatic
equilibrium in the dark-halo potential (e.g. White \& Frenk
1991; Benson et al. 2000), because of the shallower gas density 
profiles.

 It should be pointed out that the results obtained here are 
valid only for isolated late-type galaxies.
For galaxies in clusters and groups, the X-ray emission
may be affected by the intracluster (intragroup) medium.
For elliptical galaxies which contain many stars, the 
interaction between supernova-driven winds and gas 
accretion may also play an important role 
in the formation of X-ray haloes (e.g. Brighenti
\& Mathews 1999). Thus, the most relevant observational 
results to compare our model predictions with are 
those obtained by Benson et al. (2000) 
for the diffuse X-ray emission from 3 late-type galaxy 
haloes. The upper limits on the bolometric 
luminosities of these three haloes are of the order 
$10^{41}\erg\sec^{-1}$. These observational 
results are consistent with our model prediction,
if the circular velocities of these haloes (at virial
radius) are $\vcir\sim 200\kms$. If the circular velocities
of these galaxies were as high as the measured 
rotation velocities of the disks ($\sim 300\kms$, 
corresponding to a virial temperature of about 
$0.3\,{\rm keV}$), shock heating of the gas by 
gravitational accretion becomes significant.
Calculations by Tozzi \& Norman (2001) including 
this effect give an X-ray luminosity of 
$\sim 10^{41}\erg\sec^{-1}$ for haloes with 
$\Tvir=0.3\keV$, assuming a preheated
entropy similar to that which we adopt here. Thus, preheating
can help to alleviate the X-ray halo problem 
for late-type galaxies. 
  
\subsection{Accretion of cooled clouds}

\begin{figure}
\centering  
\vskip-0.5cm
\psfig{figure=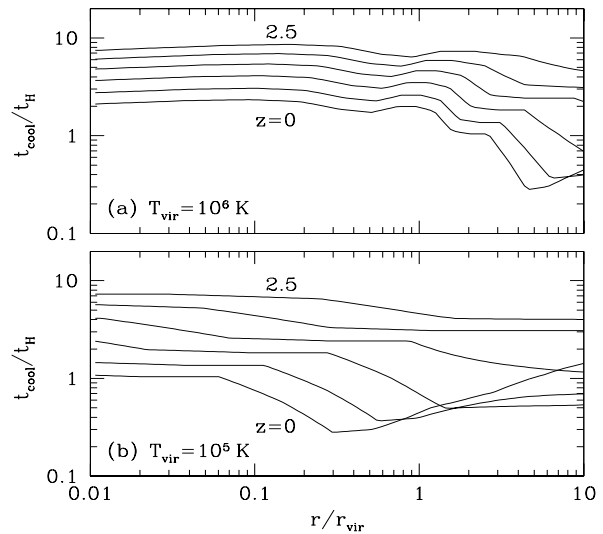,width=8.5cm,height=8.5cm,angle=0}
\vskip-0.5cm
\caption{The ratio between cooling time and Hubble
time of preheated gas around dark haloes. 
Results are shown for (a) a normal galaxy halo 
with $\Tvir=10^6\Kdegree$
and (b) a dwarf galaxy halo with $\Tvir=10^5\Kdegree$.}
\label{fig_tcool}
\end{figure}

Given the density and temperature profiles, we can estimate
the cooling time scale once the metallicity of the gas is
known. Figure \ref{fig_tcool} shows the ratio 
between the cooling time $t_{\rm cool}$ and the Hubble time
$t_{\rm H}$ assuming that the gas has a metallicity
$Z=0.1 Z_\odot$ (the cooling time is reduced by a factor
of about 2 if we assume $Z=0.3 Z_\odot$). 
We used the cooling function given in Sutherland \& Dopita (1993).
Notice that the smaller the ratio 
$t_{\rm cool}/t_{\rm H}$, the more effective the cooling is.  
Effective cooling is expected when 
$t_{\rm cool}/t_{\rm H}\sim 1$. As one can see from 
the figure, cooling is more effective at low redshift.
The main reason for this is that the gas density around
dark haloes does not change very rapidly with redshift
(and so the cooling time does not change rapidly) while
the Hubble time increases with decreasing 
redshift. The situation 
is quite different in the absence of preheating, where
the characteristic gas density 
changes with redshift as $(1+z)^3$ and
cooling is always more effective at higher redshift.
Note that once cooling becomes important, the gas can no
longer be considered adiabatic. Thus, the
adiabatic assumption may break down at low redshift.
The ratio $t_{\rm cool}/t_{\rm H}$ does not change 
significantly with radius at a given redshift, 
implying that the cooling efficiency is quite 
independent of radius. This is very different from what 
one obtains based on the assumption that 
gaseous haloes are (approximately) isothermal spheres, 
where gas cooling is always inside-out.  

Because the cooling time is quite long for the preheated
medium, most of the gas can remain in the hot phase. 
But some gas can cool due to radiative cooling. 
Since the preheated gas is expected to be clumpy, cooling 
is expected to start from regions where the gas density is enhanced, 
and a two-phase medium may develop due to thermal instability, 
with the hot phase at the temperature set by preheating 
and a cold phase at $\sim 10^4\Kdegree$ (below which 
radiative cooling may be inefficient, particularly in the 
presence of photoionization heating). The cold 
clouds produced in this way may be gravitationally 
accreted into dark haloes.

\begin{figure}
\centering  
\vskip-0.5cm
\psfig{figure=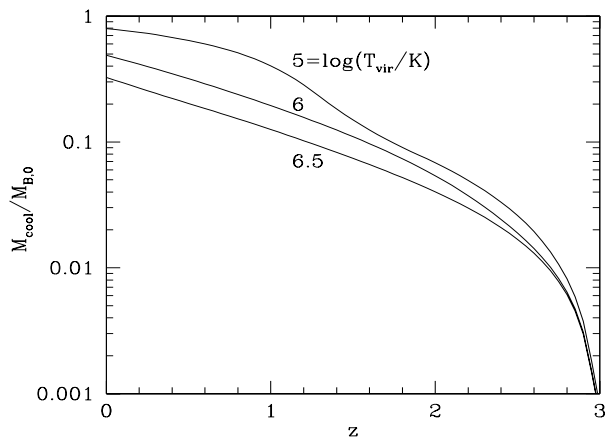,width=8.5cm,height=8.5cm,angle=0}
\vskip-1.5cm
\caption{The mass fraction in cooled gas 
(defined as the ratio between the mass of cooled gas
$M_{\rm cool}$ and $M_{\rm B,0}$, with 
$M_{\rm B,0}$ equal to $f_{\rm B}$ times the total
halo mass at the present time) as a function 
of redshift in protogalaxy regions which were preheated 
at $z=3$. The specific entropy is assumed to be $\s100=1.0$
and the gas metallicity is assumed to be $0.1 Z_\odot$. 
Results are shown for haloes with 
$\Tvir=10^5$, $10^6$ and  $10^{6.5}\Kdegree$.}
\label{fig_mcool}
\end{figure}

Since both the density and temperature of the gas 
may change with time as the dark halo grows, and since
the medium is likely inhomogenous and multi-phase, 
an accurate estimate of the amount of cold gas that can 
be accreted in a protogalaxy can be obtained
only with the use of numerical simulations. Here we give an 
approximate estimate based on simple assumptions. 
We assume that as a dark halo grows with time, 
its circular velocity $\vcir$ remains constant.  For simplicity,
we assume the dark halo mass profile is described by  a singular isothermal 
sphere, and so its radius increases with time as 
$\r200\propto 1/H(z)$ (see eq. \ref{eq:virial}).
At any given time, most of the gas is assumed to be
hot, with the gas density profile given by equations 
(\ref{deltaR1}) and (\ref{deltaR2}), while part of 
the gas in high-density regions cools.
Assuming further that the total gas mass within a 
spherical mass shell does not change with time, we can estimate 
(at any given redshift) the cooling time using the 
equilibrium density and temperature of the mass shell
in consideration. From the definition of the cooling time, 
the fraction of the gas that has cooled by redshift $z$ 
can be estimated from 
\begin{eqnarray}\label{eq_fcool}
f_{\rm cool}(z)
\equiv \int {\rmd t\over t_{\rm cool}} 
~~~~~~~~~~~~~~~~~~~~~~~~~~~~~~~~~~~~~~~~~~~~~~~&\nonumber\\
\approx
1.5\times 10^{-3} h^{-1} 
\left({\Obaryon h^2\over 0.02}\right) 
~~~~~~~~~~~~~~~~~~~~~~~~~~~~~~~~~&\nonumber\\
\times
\int_{z_{\rm ph}}^z
{H_0\,\rmd t\over \rmd z'}
{\Lambda_{-23}\over T_6} (1+\delta)
\left[1-f_{\rm cool}(z')\right] 
(1+z')^3\,\rmd z',&
\end{eqnarray}
where $z_{\rm ph}$ is the redshift at which the gas is
preheated, and the factor $[1-f_{\rm cool}(z')]$
in the integrand takes into account the depletion
of the hot gas due to cooling. Since the density 
of a fixed mass shell is known as a function of $z$
in our model, we can obtain $f_{\rm cool}(z)$ 
for each mass shell. We can then integrate over all the mass 
shells to obtain the total amount of gas that has cooled
in the protogalaxy region by some redshift $z$.    
Figure \ref{fig_mcool} shows the ratio,
$M_{\rm cool}/M_{\rm B,0}$ (where $M_{\rm B,0}$ 
is equal to $f_{\rm B}$ times the total halo mass 
at the present time) as a function of $z$.
Several important conclusions can be drawn from the figure.
(i) In all cases, significant cooling occurs at low 
redshift, for the reasons discussed earlier in this subsection.
(ii) For small haloes, rapid gas cooling 
occurs at $z\sim 1$ when the temperature of the preheated
gas reaches the level $T\la 2\times 10^5\Kdegree$
where the cooling rate peaks. In this case, the large amount of
cooled gas may trigger episodes of star formation, which
may in turn drive outflows and prevent further accretion of cooled gas.
(iii) For large haloes, the fraction 
$M_{\rm cool}/M_{\rm B,0}$ [which is roughly proportional to
$(1+z)^{-3/2}$ at low $z$] is significantly smaller than 1,
and so only a fraction of the 
gas in the protogalaxy region can cool and be accreted.

\section {Galaxy formation in a preheated intergalactic medium}
\label{sec3}

 In the last section we have shown that gas accretion 
by galaxy haloes in a preheated medium has several 
distinct properties. In this section, we discuss the 
implications of such gas accretion for the formation of 
galaxies. 

 Based on the discussion in the last section, we 
may roughly divide stars in a galaxy into two populations:  
one formed in starbursts before preheating and the other 
formed due to the accretion of gas in the preheated medium. 
The relative importance of these two populations may 
determine the morphological type of the galaxy,
because the early starbursts were likely to form
a bulge component, while an extended disk can only
form through gas accretion.

\subsection {The formation of disk galaxies}     
\label{sec3.1}

 As we have seen in Section \ref{sec2}, the gas accretion in a 
preheated medium by dark haloes occurs at rather low
redshifts (more than half of the gas is accreted 
at $z\la 1$). Thus, galaxy disks are expected to form
late in this model. In fact, late formation of 
present-day galaxy disks is required in the standard
model of disk formation (where disk size is determined by 
the specific angular momentum of dark haloes), because
the formation at high redshifts ($>1$)
predicts disks that are too compact to match the observed 
sizes (e.g. MMW). The late formation of galaxy disks
(relative to the bulges) may also be required in order to
solve the G-dwarf problem (Ostriker \& Thuan 1975).

 Another important property of the disk formation in a 
preheated medium is that only part of the gas in a 
proto-galaxy region can settle into a disk in the 
galaxy halo. Based on the discussion in Section \ref{sec2}, this
fraction is about $f=0.3$ -- 0.5 for galaxy haloes, and so the 
ratio between disk mass and halo mass is 
$m_d\sim f\Obaryon/\Omega_0 \sim 0.04$ -- 0.06 
(where we have assumed $\Obaryon=0.02 h^{-2}$, 
$\Omega_0=0.3$ and $h=0.7$). This fraction is in fact 
required in the standard model of disk formation.
Indeed, if all of the gas in a protogalaxy region 
can settle in the disk, then $m_d \sim 0.13$ in the
standard $\Lambda$CDM model. Such a high value 
of $m_d$ leads to too massive disks which have too peaked 
rotation curves and may be unstable (MMW). 
The model present here naturally leads to the
required reduction. 

\begin{figure*}
\centering  
\vskip-1.0cm
\psfig{figure=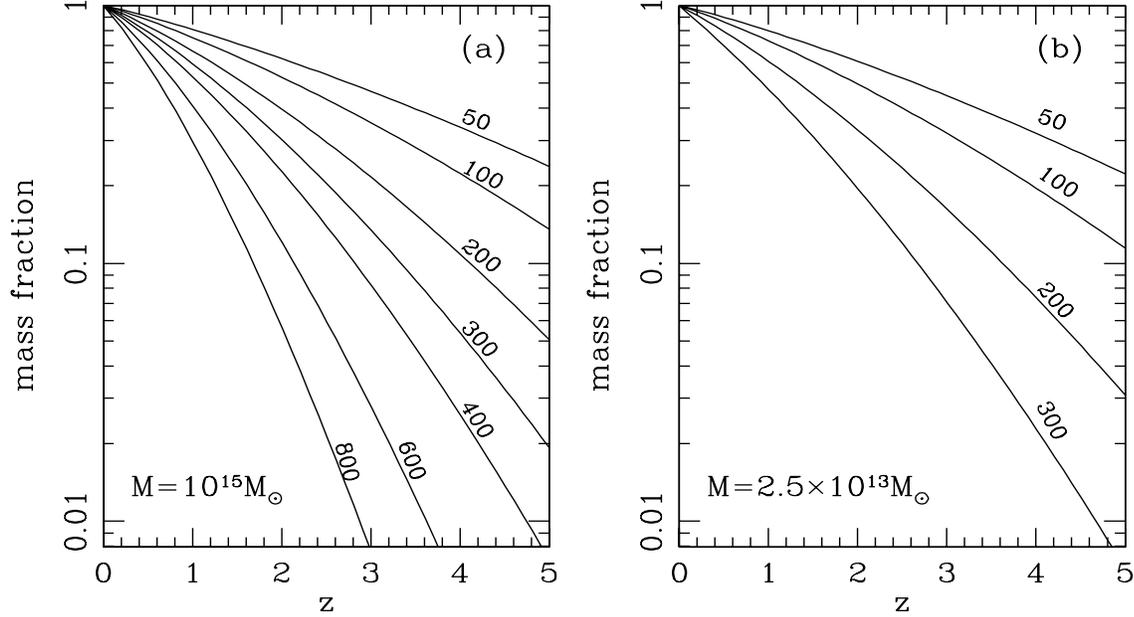,width=16.cm,height=16.cm,angle=0}
\vskip-6.0cm
\caption{The fraction of the total halo mass $M$ 
in progenitors with circular velocities exceeding 
$\vcir$ as a function of redshift. Panel (a)
shows the results for a present-day rich cluster
with mass $M=10^{15}\msun$, while panel (b)
shows the results for a present-day group
with mass $M=2.5\times 10^{13}\msun$. The circular 
velocities of progenitors are labeled on the curves
in units of $\kms$. The calculations are for the standard
$\Lambda$CDM model based on the extended Press-Schechter 
formalism (e.g. Bond et al. 1991).}
\label{fig_confraction}
\end{figure*}

 In the standard model of disk formation, the specific 
angular momentum of disk material is assumed to be 
comparable to that of dark matter. Since disk 
material and dark matter is well mixed in a 
protogalaxy, this assumption may be valid before
the collapse of the dark halo. However, in order for the 
assumption to hold for the formed disk, one has to 
make the assumption that there is no significant 
angular-momentum transfer from the disk material to 
dark matter during the formation of disks. 
As shown by the simulations of Weil et al. (1998),
this requires the gas to remain in a diffuse
form before it collapses into the final halo.
This is exactly the case in the present model, 
where most of the gas remains in a hot, diffuse form until 
late time ($z\la 1$) when the final halo collapses.
Since this gas does not suffer 
the same dynamical friction as dark matter 
clumps, the final distribution of specific
angular momentum of the gas may be different from that 
of the dark matter. 

 To see this point more clearly, we consider a 
simple case where a smaller halo, $\M200$, 
merges into a larger halo, $\M200'$. For simplicity, 
we assume the total mass profiles in both haloes are
described by that of a singular 
isothermal sphere, while the gas distribution in the small halo is given
by eq. (\ref{deltaR2}).
The smaller halo is taken to be always on a circular orbit
as it sinks towards the center of the larger one
due to dynamic friction. Because of tidal stripping, 
the outer part of the smaller halo will be 
truncated, and we estimate the truncating radius
$r_t$ on a orbit of radius $r'$ by $\rho(<r_t)=\rho'(<r')$.
For the assumed density profile, we have
$r_t=(\r200/\r200') r'$, and the remaining mass of the 
smaller halo within $r_t$ is $M(<r_t)=\vcir^2 r_t/G$. 
For material on circular orbits in the bigger halo, 
the specific angular momentum is $j(r')= \vcir' r'$. 
It is then easy to show that after merger the mass of 
dark matter that originated in $\M200$ with specific angular
momentum smaller than $j$ is
\beq\label{MjDM}
M(<j) =\M200 {j\over \jvir}\,,
\eeq
where $\jvir=\vcir' \r200'$. 
For the gas, the corresponding mass is given by
\beq
M_{\rm gas} (<j)
=\int_0^{r_j} 4\pi \rho_{\rm gas} (r) r^2\,\rmd r\,,
\eeq
where $\rho_{\rm gas}(r)$ is the gas mass density 
around the smaller halo, and $r_j=\r200 j/\jvir $.
For the gas density profile given by equation (\ref{deltaR2}),
\begin{eqnarray}
M_{\rm gas} (<j)
&=&4\pi {\overline\rho}_{\rm gas} 
\r200^3 (1+\alpha)
\nonumber\\
&&\times \int_0^{j/\jvir}
x^2\left[1-{\alpha\over 1+\alpha}\ln x\right]^{3/2}
\rmd x,
\end{eqnarray}
where $\alpha\equiv (4/5)\Tvir/({\cal S}{\overline n}^{2/3})$.
Since the logarithmic term changes only slowly with $x$, 
we have, to a good approximation, 
\beq
M_{\rm gas}(<j) \propto \left({j\over \jvir}\right)^3\,.
\eeq
Numerically, we found that for
$\alpha=0.1$, $M_{\rm gas}(<j) \propto j^{2.9}$ and for
$\alpha=10$, $M_{\rm gas}(<j) \propto j^{2.6}$. Clearly this
distribution is much steeper than that given by 
equation (\ref{MjDM}) for the dark matter. 
In fact, the shallower the gas density profile is, the steeper
the dependence of $M_{\rm gas}(<j)$ on $j$. Physically, because
of the shallow density profile, most of the gas is 
stripped in the outer part of the bigger halo and so retains 
higher specific angular momentum; only a small fraction of the gas
can sink to the inner part and lose a large amount of angular 
momentum. This effect exists wherever the gas distribution
is more extended than the dark matter.  
We emphasize that this effect remains even if we have multiple accretion
events that occur with different angles and this will lead
to a reduction in the net specific angular momentum of the accreted
gas. However, the specific angular momentum of the gas will still be
higher than that of the dark matter, if both are reduced by the
same factor.
This may help to solve the problem pointed out by 
Bullock et al. (2001) and van den Bosch
et al. (2001) that the assumption that the gas and 
dark matter in dark haloes have the same angular-momentum 
distribution leads to too concentrated disks.

\subsection {Formation of cluster galaxies and
morphological segregations}

 As one can see from Figure \ref{fig_confraction}, 
significant fraction of
the total mass of a present-day rich cluster
is already in group-sized systems with circular velocity
$\vcir\ga 400\kms$ at $z\ga 2$. These groups are expected
to be associated with high-density peaks in a protocluster 
region, and so are likely to end up in the central part 
of a cluster. Since galaxy haloes cannot have 
accreted significant amount of gas from the preheated gas 
at $z>1$ and since cooling of gas is already 
inefficient in such group haloes, we expect that 
galaxies which end up in the central part of 
clusters contain mainly old stars that have formed 
through starbursts before preheating. As the haloes
of these galaxies merge, the stellar components also merge 
due to dynamical friction to form bigger galaxies. Such mergers 
of galaxies continue until the galaxies are
incorporated into clusters in which dynamical friction
is no longer effective. We identify this to be the 
formation path for giant ellipticals in clusters. 
This formation is in fact similar to that in the 
standard model where cluster ellipticals are 
formed by mergers of galaxies (e.g. Kauffmann 1996).
The effect of preheating is to suppress the formation of new 
stars, because gas accretion was prohibited after 
preheating. Since a group halo at $z\sim 1$ contains 
many smaller haloes at the time of preheating,  
the formation of a giant elliptical is expected to 
be a result of multiple mergers of stellar systems 
formed through early starbursts, and clusters ellipticals 
are therefore expected to contain mainly old stars.
This early formation of elliptical galaxies 
(at least their stars) in gas-rich systems may be 
responsible for their tight colour-magnitude relation 
(e.g. Bower et al. 1992) and other properties 
which are distinct from present-day spiral galaxies 
(e.g. Ostriker 1980).

 As one can also see from Figure \ref{fig_confraction},
about $30\%$ of the total mass of a present-day rich 
cluster remains in small haloes ($\vcir\la 200\kms$) 
at $z\la 1$. This fraction is expected to end up in the 
outer part of the cluster.  
These haloes can accrete significant amount of gas before they are
incorporated into larger systems. The accreted 
gas may form new stars on a disk around an old bulge, 
although young stars may also be added to the bulge
either due to disk instability or due to the accretion 
of satellites containing young stellar component. 
If such systems merge into groups before being incorporated 
into the final cluster, the galaxies they contain may also 
merge, producing ellipticals with young stellar populations; 
otherwise they remain as spiral or S0 galaxies if their disks 
are not destroyed later in the cluster environment. 

 Similar discussion can be made for galaxy formation 
in smaller systems, such as present-day groups of galaxies. 
The only difference here is that, by definition, 
galaxy haloes can survive to later times
before being incorporated into group haloes with 
$\vcir\ga 300\kms$ (see Fig.\,\ref{fig_confraction}b) 
and so can accrete more gas to form disks. 
The mergers of such disks may 
also form elliptical galaxies, but unlike giant ellipticals 
in clusters, these ellipticals should contain a lot of 
younger stars. Since the merger progenitors here contain gaseous 
disks, such mergers can produce the tidal tails and 
starbursts observed at low redshifts. 

If we use the disk-to-bulge ratio to represent 
the morphological type of a galaxy, the above discussion 
suggests that the gas content, stellar age and 
local environment of galaxies should change systematically 
along the morphology sequence. Late-type
galaxies are formed in low-density regions 
where gas accretion lasts for a longer period, so they 
are gas richer, bluer and residing in lower density
than early-type galaxies. This trend is qualitatively 
consistent with observations 
(e.g. Roberts \& Haynes 1994; Dressler 1980; Postman \& Geller
1984). However, quantitatively this prediction has to be studied taking into
account of dynamical evolution and merging between galaxies.
Such a detailed calculation in a cosmological context is clearly 
needed but is beyond the scope of this paper. 

\section{Preheating the Intergalactic Medium}
\label{sec4}

 The preheating of the IGM has been discussed quite 
extensively in connection to the observed entropy excess in 
clusters and groups (Kaiser 1991; Evrard \& Henry 1991;
Ponman et al. 1999;
Tozzi \& Norman 2001 and references therein). In this 
section we discuss further some points which are closely
related to our discussion of galaxy formation.  

\subsection {Preheating by starbursts and AGNs}

 If we assume a stellar mass function (IMF) of the
standard Salpeter form [$n(M)dM \propto M^{-2.35} dM$, for
$0.1M_\odot < M<50M_\odot$]
and that each star of 
mass greater than $8\msun$ releases $10^{51} E_{51}\erg$ in 
kinetic energy in a supernova explosion, the total
energy output ($E_{\rm sn}$) is related to the total amount 
of stars formed ($M_\star$) by 
\begin{equation}
E_{\rm sn}\approx 7.5\times 10^{48} E_{51}(M_\star/\msun)
\erg\,.
\end{equation}
If a fraction $\epsilon$ of this energy is to heat 
gas with total mass $M$, each gas particle will
gain an energy (in terms of temperature): 
\begin{equation}
T_{\rm sn}
\sim 1.8\times 10^7 \epsilon E_{51} \left({M_\star\over M}\right)
\Kdegree\,.
\end{equation}
The ratio between this temperature and that given by 
equation (\ref{temp_s100}) is
\begin{equation}
{T_{\rm sn}\over T}
\sim 410 \epsilon E_{51} \left({M_\star\over M}\right)
\s100^{-1} (1+\delta)^{-2/3}(1+z)^{-2}\,.
\end{equation}
If we take $\delta=0$, $z=3$ and $\s100=1$, then
$\epsilon M_\star/M \sim 0.04$. 

\begin{figure}
\centering  
\vskip-0.5cm
\psfig{figure=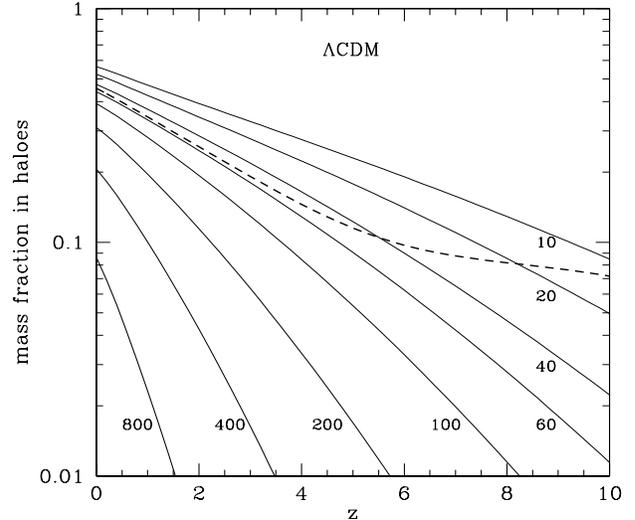,width=8.5cm,height=8.5cm,angle=0}
\vskip-0.5cm
\caption{The mass fraction of the universe in haloes
with circular velocities exceeding 
$\vcir$ (labeled on curves in units of $\kms$) 
as a function of redshift. The calculations use 
the halo mass function given in Sheth, Mo \& Tormen (2001),
and are for the standard $\Lambda$CDM cosmogony 
with $\Omega_0=0.3$, $\Omega_\Lambda=0.7$, 
$h=0.7$ and $\sigma_8=0.9$. 
The thick dashed curve
shows the mass fraction in haloes which can trap
photoionized gas [adopted from Gnedin (2000)].}
\label{fig_fraction}
\end{figure}

X-ray observations of galaxy clusters show that the mass 
in stars is about $10\%$ of the total mass in the 
X-ray gas (e.g. B\"ohringer 1995; Balogh et al. 2001). 
In this case $M_\star/M\sim 0.1$.
A similar number can be obtained as follows. The observed 
mass density in stars is $\Omega_\star\sim 2\times 10^{-3} h^{-1}$
and one-third to a half of this may be in old stellar systems, 
such as ellipticals and bulges (e.g. Fukugita, Hogan \& Peebles 1998).
If we take the standard 
$\Lambda$CDM model, the fraction of cosmic mass in
protogalaxy regions (i.e. regions which form haloes
with circular speeds $\sim 200\kms$ at the present time) 
is about $0.3$ (see Figure \ref{fig_fraction}).
Thus $M_{\rm \star, old}/M \sim 
\Omega_{\rm \star, old}/(0.3\Obaryon)\sim 0.1$.
If the heating of the IGM were all due to star formation,
an efficiency $\epsilon \sim 0.4$ would be required.
This efficiency is higher than usually assumed
(e.g. Valageas \& Silk 1999; Wu, Fabian \& Nulsen 2000;
Kravtsov \& Yepes 2000). 
However, an efficiency as high as this is in fact indicated 
by observations of supernova-driven winds in local 
starburst galaxies (Heckman et al. 2000).
Thus, if most stars at high redshifts were formed in 
systems reminiscent of local starbursts, as suggested by the observed
properties (high star-formation rate, compact size)
of high-redshift galaxies (see Heckman 2001 and references
therein), the stellar energy sources may be sufficient to preheat 
the IGM to the required level. More quantitatively, the observed 
mass-outflow rate in a local starburst is usually comparable 
to the star-formation rate, and the wind velocity is 
typically $500$ -- $600\kms$. If the kinetic energy
of the wind is to be thermalized in the IGM, it can heat as 
much as ten times $M_{\rm wind}$ (the mass of the wind) 
of the IGM to the required temperature at $z\sim 3$
[see equation (\ref{temp_s100})]. Since 
$M_{\rm wind}\sim M_\star$, the total preheated mass 
is $M\sim 10 M_\star$, consistent with the star/gas 
mass ratio estimated above.     
  
An alternative energy source for the preheating may 
be AGNs. For example, radio galaxies can provide 
a total energy output about two orders of magnitude 
larger than stellar sources, and so have the potential
to heat the IGM to the required level (e.g. 
Inoue \& Sasaki 2001). Since the mass of central 
black hole (which powers the AGN) is found to be 
proportional to the bulge mass in a galaxy 
(e.g. Magorrian et al. 1998; Gebhardt et al. 2000), the energy 
output from AGNs in a region large enough to contain many 
AGNs should be roughly proportional to the energy output 
from stellar sources. Thus, the 
inclusion of additional energy output from AGNs is equivalent 
to increasing the value of $\epsilon$. The uncertainty here
is again how large a fraction of the total energy
released by AGNs can be thermalized in the IGM.   

\subsection{The epoch of preheating}    

 If the preheating was indeed done by activity associated
with star formation and AGNs, the likely epoch for preheating
is when the star-formation and AGN activity peaks.
The observed comoving number density of AGNs peaks 
around $z=2$ to 3 (Shaver et al. 1996). The star formation 
history in the universe can be inferred from the observed 
${\dot\Omega}_\star (z)$ which is the change rate of the 
density parameter of stars. Observations showed that
${\dot\Omega}_\star$ remains approximately constant
($\sim 10^{-3}{\,\rm Gyr}^{-1}$ ) in the redshift range 
$z=2$ - 4, and decreases rapidly with decreasing $z$ at $z<1$
(e.g. Blain et al. 1999).
For the $\Lambda$CDM model, the 
age of the universe at $z\sim 3$ is about 1.5 Gyr, 
and so the mass density parameter of all stars formed before $z=3$ 
is $\Omega_\star \sim 10^{-3}$, consistent with the 
total amount of stars observed in old stellar systems.
Thus, if the mass 
of gas that can be heated up to the required entropy
is about ten times that of the old stars, the density
parameter of the heated gas is 
$\Omega_{\rm gas}\sim 0.01\sim 0.2\Obaryon$, which
is similar to the gas mass in proto-group regions
(see Fig.\,\ref{fig_fraction}).
 From these considerations, we see that the 
epoch of preheating is likely to be around $z=2$ to 3.

 Similar conclusion can be drawn from theoretical 
considerations. As shown in Figure \ref{fig_confraction}a, 
about $20\%$ of the total mass of a present-day cluster is already in 
progenitors with circular velocities $\vcir\ga 200\kms$
at $z\sim 3$. This fraction is only slightly smaller for
a present-day group (see Fig.\,\ref{fig_confraction}b). 
Most systems that have formed by redshift 3
will suffer a major merger in a Hubble time
(see Fig.14 in Shu, Mao \& Mo 2001), and so most 
of them may form stars in starbursts. If half of the gas in 
these systems forms stars, and the other half escapes as outflows with
the escaping velocity, 
$v_{\rm esc}\sim 3 \vcir$ [where we assumed a singular isothermal
sphere and the gas escapes at roughly one disk scale length, cf.
equation (12) in MMW and equation (2-192) in Binney \& Tremaine
(1987)], then the IGM in the protogroup
(or protocluster) regions can be preheated to the
required level at $z\sim 3$. As one can see from 
Fig \ref{fig_confraction}, about half of the total 
mass of a present-day cluster is in progenitors with 
circular velocities $\vcir\la 50\kms$ at $z\sim 3$.
Since such haloes cannot trap much gas in the presence
of the general UV background (e.g. Efstathiou 1992;
Gnedin 2000), more than half of the 
protocluster (protogroup) gas must be in diffuse form. Thus, 
most of the gas to be heated is in a diffuse medium
outside collapsed haloes. The situation here is different 
from that in previous feedback schemes without preheating, where 
the goal is to heat up and get rid of the gas that has 
already been assembled into galaxies or galactic haloes.

In reality, the time of preheating must be different 
in different regions. In high-density regions where
a significant fraction of the mass was already in 
galactic haloes at high redshifts, the time of 
preheating is expected to be early, while in 
low-density regions where galaxy haloes form at low redshifts, 
the time of preheating is expected to be late. 
In regions where starbursts never occur, 
the IGM is not preheated. A detailed modelling of how and 
when different regions in the universe were heated up
depends on how the heating sources form in the cosmic 
density field and how their energy is thermalized in the
IGM. This is a complicated problem and will not be 
considered in this paper. 
 
\subsection {Preserving the ${\rm Ly}\alpha$ forest} 

If all of the IGM were heated to the temperature implied
above, then there would be no photoionized gas to produce 
the observed ${\rm Ly}\alpha$ forest. This is not allowed, because
the ${\rm Ly}\alpha$ forest is observed. However, the above discussion
does not require the preheating to be everywhere in the IGM. 
In fact, in order to explain the observed 
entropy excess, it is only necessary to preheat the gas which
ends up in present-day groups and clusters. Since groups and clusters are
places where vigorous star-formation and AGN activities are expected
in the early time, it is also {\it likely} that preheating occurred 
mainly in protocluster (protogroup) regions. In fact, 
star-forming galaxies at high redshifts, such as  
Lyman-break galaxies and sub-mm sources observed at 
$z\sim 3$, are expected to be located in high density regions which 
will eventually be in present-day groups and clusters
(e.g. Mo, Mao \& White 1999, and references therein).  
It is therefore possible that the IGM in the low density regions,
which may be responsible for the observed ${\rm Ly}\alpha$ forest, 
is not affected significantly, while the high density regions 
are heated. To demonstrate that this is likely the case, 
we can estimate the fraction of the total mass which is in 
systems more massive than present-day groups 
(i.e. with circular speeds $\ga 400\kms$).
As shown in Figure \ref{fig_fraction},
this fraction is about $20\%$ in standard $\Lambda$CDM.
In fact more than half of the total mass is in systems which 
are too small to trap photoionized gas. These systems 
are outside protogalaxy regions, and so may not be influenced 
significantly if preheating is confined to proto-group 
regions by the associated gravitational potentials.

Of course, preheating may produce some effects that can be observed 
in the ${\rm Ly}\alpha$ forest (Theuns, Mo \& Shaye 2001).
If the preheating is due to supernova-driven winds from 
starbursts, some of the metals ejected by the heating sources 
may leak to the Lyman forest and contaminate it, although 
most of the metals may be hidden in the hot phase.
This may in fact be the origin of the `missing metal' 
problem that the total amount of metals observed 
at $z\sim 3$ seems smaller than that expected from 
the observed star formation history (e.g. Pettini 1999).

\section {Discussion and Summary}
\label{sec5}

   In this paper, we have outlined a scenario of star-formation 
feedback in which the IGM in galaxy-forming regions was preheated. 
The preheating of the intergalactic medium in high-density regions 
of the cosmic density field may be responsible for the entropy excess 
observed in low-mass clusters of galaxies and yet may not destroy 
the bulk of the ${\rm Ly}\alpha$ forest. 
The subsequent galaxy formation in preheated media
is different from that in models without preheating,
because the gas component no longer follows the bottom-up 
clustering hierarchy of the dark matter component 
on galaxy scales. Because of the initial entropy,
the hot gas around a galaxy halo has a very shallow profile, and 
so emits only weakly in X-ray. This may be the reason 
why extended X-ray haloes are not yet detected  
around spiral galaxies. Radiative cooling of the 
preheated gas around galaxies may become more significant 
at lower redshift when the cosmic time becomes longer. 
Unlike in the absence of preheating, where cooling is always 
inside-out in a halo, the cooling efficiency in a preheated 
halo does not change rapidly with radius. 
Because of thermal instability, the cooled gas may form cold 
clouds which can then sink toward the centre 
in the potential wells of galaxy haloes 
to form galaxy disks. The total amount of the gas that a galaxy halo can 
accrete by the present time is significantly smaller than the total gas 
in the protogalaxy region. Since most gas is accreted 
in diffuse clouds, the gas may not lose angular momentum to 
dark matter due to dynamical friction. Cluster ellipticals 
are produced by the mergers of stellar 
systems formed prior to the preheating, because these systems 
cannot accrete much gas from the preheated medium to form 
new stars before they are incorporated into large haloes where 
gas cooling is inefficient. Large galaxy disks form in low-density 
environments where accretion of gas can continue to the 
present time. Mergers of such disks may trigger new starbursts
and form field ellipticals. Thus, the stellar age, gas content, 
and local environment of galaxies are expected to change 
systematically along the morphology sequence. 

  The scenario proposed here can be tested by observations
of its assumptions and predictions. If part of the IGM 
was indeed heated up to a high temperature, we may hope to 
find some signatures of such heating in the $\Lya$ forest
(Theuns, Mo \& Schaye 2001). Since the preheating is most 
likely in regions of vigorous star formation, observations of 
the $\Lya$ forest near star-forming galaxies, such as 
the Lyman-break population (e.g. Steidel et al. 1999), 
are important for probing the 
properties of the preheating. Because the preheated gas 
contains metals and is at a temperature much higher than
$10^4\Kdegree$, observations of absorption lines of 
collisionally ionized metal species (e.g. O VI) are also important.  
Another test may come from observations of the diffuse
X-ray emission from spiral-galaxy haloes. As discussed in 
Section \ref{sec2}, current observational results are still too 
uncertain to provide stringent constraints. Future 
observations from the more sensitive detectors, such as 
XMM and Chandra may be used to probe the gaseous haloes
predicted here. 

Yet another test of the model may come from
observations of the cold gas around galaxy haloes. An 
important probe here is provided by QSO absorption line
systems associated with low-redshift galaxies. 
Such systems have been observed. For example, normal galaxies 
at moderate redshifts ($z\la 1$) are observed to produce
strong MgII systems whenever the impact parameter 
to a QSO sightline is smaller than $\sim 40h^{-1}\kpc$
(Steidel 1995 and references therein), suggesting that 
each galaxy may possess a halo of cold clouds.
Weaker MgII systems are also
observed to be associated with galaxies at larger 
impact parameters (e.g. Rigby, Charlton \& 
Churchill 2002). These absorption systems may be produced by
the cooling clouds predicted in the present model.
Indeed, there are theoretical investigations along 
this line (e.g. Mo 1994; Mo \& Miralda-Escude 1996; 
Lin et al. 2000). Assuming that strong MgII systems 
are produced by gas clouds cooling from and 
pressure-confined by hot haloes similar to those 
considered here, Mo \& Miralda-Escude found that many of the
observed properties of the absorber/galaxy systems can be 
reproduced. Clearly, more observations of this kind  
can give stringent constraints on the gaseous galaxy 
haloes we are proposing here. 

 On the theoretical side, there are several important 
issues for which further investigations are required. 
For the preheating, we need a physical model to predict
how the IGM was heated. For simplicity, we have modelled
the preheated IGM as a uniform medium. This
is clearly a simplification, as the density contrast is likely
a smooth function of radius, decreasing from the center
of the preheating energy sources (bulges or AGNs)
to the preheated IGM. Also if the preheating is accompanied
by outflows, the interplay between outflows and gas
accretion needs to be modelled in detail. Some of these issues 
are best addressed using numerical simulations. Finally,
we also need to incorporate the 
present scenario of galaxy formation into realistic
cosmological models to make detailed predictions 
for the properties of the galaxy population. We intend
to return to some of these issues in future papers.

\section*{Acknowledgement}

We thank James Bullock, Jerry Ostriker, Frank van den Bosch, Simon White 
and the refereree, Peter Thomas, for helpful comments and suggestions.

\end{document}